\documentstyle[12pt,epsfig]{article}
\input{epsf}

\def\la{\mathrel{\mathpalette\fun <}}

\def\fun#1#2{\lower3.6pt\vbox{\baselineskip0pt\lineskip.9pt
\ialign{$\mathsurround=0pt#1\hfil##\hfil$\crcr#2\crcr\sim\crcr}}}

\begin{document}
\title{Systematization of tensor mesons and the determination of the
$2^{++}$ glueball}

\author{V. V. Anisovich \footnote{E-mail: anisovic@pnpi.spb.ru}
\\
Petersburg Nuclear Physics Institute, Gatchina,
188300, Russia}
\date{}
\maketitle

\begin{abstract}

It is shown that new data on
the $(J^{PC}=2^{++})$-resonances in the mass range
 $M\sim1700-2400\,$MeV support the linearity of the
$(n,M^2)$-trajectories, where  $n$ is the radial quantum number of
quark--antiquark state. In this way all vacancies for
the isoscalar tensor $q\bar q$-mesons
in the range up to 2450 MeV are filled in.
This allows one to fix the broad  $f_2$-state with  $M=2000\pm30\,$MeV
and $\Gamma=530\pm40\,$MeV as the lowest tensor glueball.  \\

PACS numbers: 14.40.-n, 12.38.-t, 12.39.-Mk

\end{abstract}

Recent analysis of the
process $\gamma\gamma\to K_SK_S$ \cite{L3} and re-analysis
of $\phi\phi$-spectra \cite{LL} observed in the reaction
$\pi^-p\to\phi\phi n$ \cite{Etk} have clarified the
situation with $f_2$-mesons in the mass region $1700-2400\,$MeV.
Hence, now one may definitely speak about the location  of
$q\bar q$-states on the $(n,M^2)$-trajectories \cite{syst}, see also
\cite{book,ufn04}. This fact enables us to determine which one of
$f_2$-mesons is an extra state for the $(n,M^2)$-trajectories. Such
an extra state is the broad resonance $f_2(2000\pm30)$. According to
 \cite{LL,Bar,Ani}, its parameters are as follows:
\begin{eqnarray}
&&M=2050\pm30\mbox{ MeV },\quad \Gamma=570\pm70\mbox{ MeV } \cite{LL},
\nonumber\\ &&M=1980\pm20\mbox{ MeV },\quad \Gamma=520\pm50\mbox{ MeV }
\cite{Bar}, \nonumber\\
&&M=2010\pm25\mbox{ MeV },\quad \Gamma=495\pm35\mbox{ MeV } \cite{Ani}.
\label{1}
\end{eqnarray}
In \cite{syst}, we have put quark--antiquark meson states with different
radial excitations  $(n=1,2,3,4,\ldots)$ on the
$(n,M^2)$-trajectories. With a good accuracy, the trajectories
occurred to be linear:
\begin{equation} M^2\ =\
M^2_0+(n-1)\mu^2 \ ,
\label{2}
\end{equation}
with a universal slope
$\mu^2=1.2\pm0.1\,\rm GeV^2$; $M_0$ is the mass of the
lowest (basic) state.
For the  $(I=0, J^{PC}=2^{++})$-mesons, the present status of
trajectories (i.e. with the results given by \cite{L3,LL}) are shown in
Fig. 1.

The quark states with $(I=0, J^{PC}=2^{++})$ are defined by two flavour
components, $n\bar n=(u\bar u+d\bar d)/\sqrt2$ and $s\bar s$,
with $^{2S+1}L_J=\,^3P_2, ^3F_2$. Generally, all mesons are the
mixture of flavour component in the $P$- and $F$-waves. But, as concern
the  $f_2$-mesons with $M\la2\,$GeV, they are dominated by the flavour
component $n\bar n$ or $s\bar s$ in a definite
$P$ or $F$ wave. The $f_2$-mesons shown in Fig. 1, which belong to four
trajectories, are dominated by the following states:
\begin{eqnarray}
\bigg[f_2(1275),f_2(1580), f_2(1920), f_2(2240)\bigg]&\longrightarrow&
^3P_2n\bar n\ , \nonumber\\
\bigg[f_2(1525),f_2(1755), f_2(2120), f_2(2410)\bigg]&\longrightarrow&
^3P_2s\bar s\ ,
\nonumber\\
\bigg[f_2(2020),\ f_2(2300)\bigg] &\longrightarrow& ^3F_2n\bar n\ ,
\nonumber\\
f_2(2340) & \longrightarrow& ^3F_2 s\bar s\ .
\label{3}
\end{eqnarray}
To avoid the confusion, in (\ref{3}) the experimentally  observed
masses of mesons are shown  --- these are the
magnitudes drawn in Fig. 1 but not those from the compilation
\cite{PDG}.

 Let us discuss the states which lie on the trajectories of Fig. 1.

\subsection*{The trajectory $\bf[f_2(1275),f_2(1580), f_2(1920),
f_2(2240)]$}

1) $f_2(1275)$: This resonance is almost pure
$1^3P_2n\bar n$-state:
this is favoured by the comparison of branching ratios
$f_2(1275)\to\pi\pi,\eta\eta,K\bar K$ with quark model calculations.
The dominance  of $1^3P_2n\bar n$ component is
also supported by  the value of partial width of the decay
 $f_2(1275)\to\gamma\gamma$ \cite{tensorYF,epja}.

\noindent2) $f_2(1580)$ (in the compilation \cite{PDG} it is denoted as
$f_2(1565))$: About ten years ago, there existed a number of
indications to the presence of the  $2^{++}$-mesons in the
vicinity of 1500~MeV \cite{Aker,Adamo,bertin,Armstrong}. After the
discovery of a strong signal in the $0^{++}$-wave related to the
$f_0(1500)$ \cite{AniPL,AniPR} as well as  correct account for the
interference of $0^{++}$ and $2^{++}$
waves, the resonance signal in the $2^{++}$ wave
moved towards higher masses, $\sim1570\,$MeV. According to the
latest combined analysis of meson spectra
 \cite{ufn04,PNPI}, this resonance has the following characteristics
(see  Table~1 in \cite{ufn04}):
\begin{equation} M=1580\pm6\mbox{ MeV
}, \quad \Gamma=160\pm20\mbox{ MeV }\ .
\label{4}
\end{equation}
Hadronic decays together with partial width in the channel
$\gamma\gamma$ \cite{tensorYF} support the
$f_2(1580)$ as a system with dominant $n\bar n$-component.

In \cite{PDG}, the  $f_2(1640)$-state is marked as a separate
resonance: this identification is based on resonance signals at
$M=1620\pm16\,$MeV
\cite{Bugg} (Mark 3 data for $J/\Psi\to\gamma\pi^+\pi^-\pi^+\pi^-$),
$M=1647\pm7\,$MeV \cite{Adamo} (reaction $\bar np\to3\pi^+2\pi^-)$,
$M=1590\pm30\,$MeV \cite{VES-f2}, $1635\pm7\,$MeV \cite{GAMS-f2}
(reaction $\pi^-p\to\omega\omega n$).
Without doubt, these signals are the reflections of $f_2(1580\pm20)$,
and the data  \cite{Bugg,VES-f2} do not contradict this fact.
In \cite{PDG}, the mass of this state is determined as
$1638\pm6\,$MeV that reflects small errors in the mass
definition in  \cite{Adamo,GAMS-f2}.

\noindent3) $f_2(1920)$ (in the compilation \cite{PDG}, it is
denoted as  $f_2(1910)$):  This resonance was observed in the signals
$\omega\omega$ \cite{VES-f2,GAMS-f2,WA} and $\eta\eta'$
\cite{GAMS-eta,WA-eta}. In \cite{Ani}, the $f_2(1920)$ is seen as a
shoulder in the
$p\bar p(I=0,C=+1)\to\pi^0\pi^0,\eta\eta,\eta\eta'$ spectra, in the
wave $^3P_2p\bar p$.  According to \cite{ufn04,PNPI},
\begin{equation}
\label{5} M=1920\pm40\mbox{ MeV }, \quad \Gamma=260\pm40\mbox{ MeV }.
\end{equation}
A strong signal in the channels with nonstrange mesons surmises a large
 $n\bar n$ component in the $f_2(1920)$.

\noindent4) $f_2(2240)$: It is seen in the spectra $p\bar
p(I=0,C=+1)\to\pi^0\pi^0,\eta\eta,\eta\eta'$, in the wave $^3P_2p\bar p$
\cite{Ani}.  According to \cite{ufn04,PNPI}:
\begin{equation}
\label{6}
M=2240\pm30\mbox{ MeV }, \quad \Gamma=245\pm45\mbox{ MeV }.
\end{equation}
The decay of $f_2(2240)$ into channels with nonstrange mesons makes it
verisimilar the assumption about a considerable $n\bar n$ component.

\noindent5) The next radial excitation on the $^3P_2n\bar n$ trajectory
($n=5$) is predicted at 2490~MeV.

\subsection*{The trajectory $\bf[f_2(1525), f_2(1755), f_2(2120),
f_2(2410)]$}

This is the meson trajectory with dominant  $s\bar s$-component.
The states lying on this trajectory are the nonet partners of mesons
from the first trajectory [$f_2(1275)$, $f_2(1580)$,
$f_2(1920)$, $f_2(2240)$]. This suggests a dominance of the $P$-wave in
these $q\bar q$-systems: $^3P_2q\bar q$.

1) $f_2(1525)$: This is the basic state, $(n=1)$, the nonet partner of
$f_2(1275)$. The mixing angle of $n\bar n$ and $s\bar s$ components,
which can be determined  neglecting  the gluonium admixture,
\begin{eqnarray}
&& f_2(1275)\ =\ n\bar n\cos\varphi_{n=1}+s\bar s\sin\varphi_{n=1}\,,
\nonumber\\
\label{7}
&& f_2(1525)\ =\ -n\bar n\sin\varphi_{n=1}+s\bar s\cos\varphi_{n=1}\,,
\end{eqnarray}
may be evaluated from the value of the partial widths $\gamma\gamma$ and
ratios of the decay channels $\pi\pi$, $K\bar K$,
$\eta\eta$ within the frame of quark combinatorics (see \cite{book},
Chapter 5 and references therein).
Evaluations given in \cite{L3,tensorYF} provide us the mixing angle as
follows:
\begin{equation} \label{8} \varphi_{n=1}\ =\
-1^\circ\pm3^\circ\ .
\end{equation}

2) $f_2(1755)$: This state belongs to the nonet of the first radial
excitation,
$n=2$, it is dominantly the $P$-wave $s\bar s$ state. The mixing angle
$\varphi_{n=2}$ can be evaluated using the data on $\gamma\gamma\to
K_SK_S$.  Neglecting a possible admixture of the glueball component, it
was found \cite{L3}:
\begin{eqnarray}
&& f_2(1580)\ =\ n\bar n\cos\varphi_{n=2}+s\bar s\sin\varphi_{n=2}\ ,
\nonumber\\
&& f_2(1755)\ =\ -n\bar n\sin\varphi_{n=2}+s\bar s\cos\varphi_{n=2}\ ,
\nonumber\\
\label{9}
&& \varphi_{n=2}\ =\ -10^\circ\
{}^{\displaystyle{+5^\circ}}_{\displaystyle{-10^\circ}}\ .
\end{eqnarray}

3) $f_2(2120)$: This resonance was observed in the
$\phi\phi$ spectrum in the reaction
 $\pi^-p\to n\phi\phi$ \cite{Etk}.  At small momenta transferred to
the nucleon the pion exchange dominates, so we have the transition
 $\pi\pi\to\phi\phi$. The $f_2(2120)$ resonance is seen in
the $\phi\phi$ system in the
$S$-wave with the spin 2 (the state $S_2$).
According to \cite{LL}, its parameters are as follows:
\begin{equation}
M=2120\pm30\mbox{ MeV }, \quad \Gamma=290\pm60\mbox{ MeV }, \quad
W(S_2)\simeq90\%\ ,
\label{10}
\end{equation}
where $W(S_2)$ is the probability of the $S_2$-wave.
The previous analysis \cite{Etk}, that did not account for the
existence of the broad $f_2$-state around 2000~MeV, provided one
the value
$M\simeq2010\,$MeV, $\Gamma\simeq200\,$MeV \cite{Etk}, accordingly,
this resonance was denoted
 as $f_2(2010)$ in \cite{PDG}.  At the same time, there is
a resonance denoted in \cite{PDG} as
 $f_2(2150)$, which was observed in the spectra
$\eta\eta$, $\eta\eta'$, $K\bar K$, that assumes a large
$s\bar s$-component:
\begin{equation} \label{11a}
\begin{tabular}{lllr}
$\eta\eta$ \cite{Bar-etaeta}: & $M=2151\pm16$ MeV, & $\Gamma=280\pm70$
MeV, & \\
$\eta\eta$ \cite{sing}: & $\qquad2130\pm35$ MeV, & $\Gamma=130\pm30$
MeV, &\\

$\eta\eta,\eta\eta'$ \cite{Ani-etaeta}: & $\qquad2105\pm10$
MeV, & $\Gamma=200\pm25$ MeV,&\\

$\eta\eta$ \cite{Armstrong}: & $\qquad2104\pm20$ MeV, &
$\Gamma=203\pm10$ MeV, &\\

$K\bar K$ \cite{Bar-KK}: & $\qquad2130\pm35$
MeV, & $\Gamma=270\pm50$ MeV. &
\end{tabular}
\end{equation}
The re-analysis \cite{LL} points definitely to the fact that the
resonances denoted in \cite{PDG} as $f_2(2010)$ and $f_2(2150)$ are
actually the same state.

4) $f_2(2410)$: It is seen in the reaction $\pi^-p\to n\phi\phi$
\cite{Etk}.  According to the re-analysis \cite{LL},
its parameters are as follows:
\begin{eqnarray}
&& M=2410\pm30\mbox{ MeV }, \quad \Gamma=360\pm70\mbox{ MeV },
\nonumber\\
\label{11}
&&W(S_2)\simeq50\%\,, \quad W(D_0)\simeq20\%\,, \quad
W(D_2)\simeq30\%\,.
\end{eqnarray}
If the contribution of the broad
$f_2$-state in the region 2000~MeV is neglected,
the resonance parameters move to smaller values:
$M\simeq2340\,$MeV,
$\Gamma\simeq320\,$MeV \cite{Etk}; correspondingly, in \cite{PDG} it
was denoted as $f_2(2340)$.

5) The linearity of the $(n,M^2)$ trajectory predicts the next
$^3P_2s\bar s$ state at 2630~MeV $(n=5)$.

\subsection*{The states with dominant  $^3F_2n\bar n$
component}

At the time being we may speak about the observation of the two
 states with the dominant
$^3F_2n\bar n$-component.

1) $f_2(2020)$: It is seen in the partial wave analysis of
the reactions $p\bar
p\to\pi^0\pi^0,\eta\eta,\eta\eta'$, in the wave $^3F_2p\bar p$
\cite{Ani}. According to \cite{ufn04,PNPI}, its parameters are as
follows:
\begin{equation}
M=2020\pm30\mbox{ MeV }, \quad
\Gamma=275\pm35\mbox{ MeV }.
\label{12}
\end{equation} In \cite{PDG}, this meson was placed to the Section
"Other light mesons", it is denoted as
$f_2(2000)$. This is the basic  $^3F_2$-meson ($n=1)$ with the dominant
 $n\bar n$-component.

2) $f_2(2300)$: It is seen in the partial wave analysis of the
reaction $p\bar
p\to\pi^0\pi^0,\eta\eta,\eta\eta'$, in the wave $^3F_2p\bar p$
\cite{Ani}.  According to \cite{ufn04,PNPI}, its parameters are
as follows:
\begin{equation}
\label{16} M=2300\pm35\mbox{ MeV }, \quad
\Gamma=290\pm50\mbox{ MeV }.
\end{equation}
This is the first radial excitation of the
 $^3F_2$-state $(n=2)$, with dominant $n\bar
n$-component. There is a resonance denoted in \cite{PDG} as
$f_2(2300)$, but this is the state observed in the
$\phi\phi$-spectrum \cite{Etk}, the mass and width of which, in
accordance with the re-analysis \cite{LL}, are  $2340\pm15\,$MeV and
$ 150\pm 50$ MeV --- of course, they are different states, see the
discussion below.

3) The second radial excitation state ($n=3$) on the trajectory
$^3F_2n\bar n$ is predicted to be at
$M\simeq2550\,$MeV.

\subsection*{The state with dominant  $^3F_2s\bar s$ component}

This trajectory is marked by one observed  state only.

1) $f_2(2340)$: It is seen in the $\phi\phi$-spectrum \cite{Etk} and
$\gamma\gamma\to K^+K^-$ \cite{Abe}, with the mass
$\sim2330\,$MeV and width $275\pm 60$ MeV. According to \cite{LL},
\begin{eqnarray} &&
M=2340\pm15\mbox{ MeV }, \quad \Gamma=150\pm50\mbox{ MeV }, \nonumber\\
\label{17} && W(S_2)\simeq10\%\,, \quad W(D_0)\simeq10\%\,, \quad
W(D_2)\simeq80\%\ .
\end{eqnarray}
In the previous analysis of the
$\phi\phi$-spectrum \cite{Etk}, this resonance had the mass
 2300~MeV, in \cite{PDG} it is denoted as $f_2(2300)$.

2) The next state on the $^3F_2s\bar s$ trajectory ($n=2)$
should be located near $M\simeq2575\,$MeV.

\subsection*{The broad $2^{++}$-state near  2000~MeV ---
the tensor glueball}

The averaging over parameters of the broad resonance using the data
\cite{LL,Bar,Ani}, see (1), gives us the following values:
\begin{equation} \label{18}
M=2000\pm30\mbox{ MeV }, \quad \Gamma=530\pm40\mbox{ MeV }.
\end{equation}
This broad state is superfluous with respect to
$q\bar q$-trajectories on the
$(n,M^2)$-plane, i.e. it is the exotics. It is reasonable to believe
that this is the lowest tensor glueball. This statement is favoured
by the estimates of  parameters of the pomeron trajectory (e.g. see
\cite{book}, Chapter 5.4, and references therein), according to which
$M_{2^{++}glueball}\simeq1.7-2.5\,$GeV.  Lattice calculations result in
a close value, namely,  $2.2-2.4\,$GeV
\cite{2glueball}.

Another characteristic signature of the glueball is its large width,
that was specially underlined in \cite{PR-exotic}. The matter is that
exotic state accumulates the widths of its neighbours--resonances
due to the transitions  $meson(1)\to real\,mesons\to
meson(2)$.

Just this phenomenon took place with the lightest scalar
glueball near 1500~MeV:  the decay processes led to a strong mixing of
the glueball with neighbouring resonances, so the glueball
turned into the broad resonance $f_0(1200-1600)$
\cite{APS-PL,APS-ZP,AAS-PL,AAS-ZP},
see also the discussion in \cite{ufn04}. Of course, the
width of this scalar isoscalar state  is rather large, though its
precise value is poorly determined:
$\Gamma\simeq500-1500\,$MeV.  Although the accuracy in the
determination of absulute value is low, the ratios of partial widths of
this state to channels  $\pi\pi,K\bar
K,\eta\eta,\eta\eta'$ are well defined \cite{kmat}. So the ratios
$\Gamma(\pi\pi):\Gamma(K\bar K):\Gamma(\eta\eta):\Gamma(\eta\eta')$
tell us definitely that  $f_0(1200-1600)$ is a mixture of the gluonium
$(gg)$ and quarkonium  $(q\bar q)$ components being close to the flavour
singlet $(q\bar q)_{glueball}$.  Namely,
\begin{eqnarray}
\label{G1} && gg\cos\gamma+(q\bar q)_{glueball}\sin\gamma\ , \\
 && (q\bar q)_{glueball}=\ n\bar n\cos\varphi_{glueball}+s\bar s\sin
\varphi_{glueball} \nonumber
\end{eqnarray}
with
$\varphi_{glueball}=\arctan\sqrt{\lambda/2}\simeq26^\circ-33^\circ$.
The mixing angle  $\varphi_{glueball}$ is determined by the fact that
the gluon field creates the light quark pairs with probabilities
$u\bar u:d\bar d:s\bar s=1:1:\lambda$,
and the probability to produce strange quarks
$(\lambda)$ is suppressed $\lambda\simeq0.5-0.85$ (see
\cite{klempt} and the discussion in Chapter 5 of \cite{book}).  The
mixing angle $\gamma$ for gluonium and quarkonium components
cannot be defined by the ratios      $\Gamma(\pi\pi):\Gamma(K\bar
K):\Gamma(\eta\eta):\Gamma(\eta\eta')$ --- it should be fixed by
radiative transitions, for example, $\gamma\gamma \to f_0(1200-1600)$;
such an experimental information is still missing. One
may find a detailed discussion of the situation in
\cite{book,ufn04}.

If the broad resonance  $f_2(2000)$ is the tensor glueball, it must be
also the mixture of components $gg$ and $(q\bar q)_{glueball}$. Then the
decay vertices of  $f_2(2000)\to\pi\pi,K\bar
K,\eta\eta,\eta\eta',\eta'\eta'$ ¨
$f_2(2000)\to\omega\omega,\rho\rho,K^*K^*,\phi\phi,\phi\omega$ should
obey the constraints shown in  Table.

\begin{table}
\caption{The constants of the tensor glueball decay into two mesons in
the leading (planar diagrams) and next-to-leading (non-planar
diagrams) terms of $1/N$-expansion. Mixing angles for $\eta-\eta'$ and
$\omega-\phi$ mesons are defined as follows: $\eta=n\bar
n\cos\theta-s\bar s\sin\theta$, $\eta'=n\bar n\sin\theta+s\bar
s\cos\theta$ and $\omega=n\bar n\cos\varphi_V-s\bar s\sin\varphi_V$,
$\phi=n\bar n\sin\varphi_V+s\bar s\cos\varphi_V$. Because of the small
value of $\varphi_V$, we keep in the Table the terms of the order
of $\varphi_V$ only.}

\begin{tabular}{||c|c|c|c||}

\hline\hline

 & Constants for & Constants for & Identity factor \\
 & glueball decays in & glueball decays in & for decay \\
Channel & the leading order & next-to-leading order & products \\
 & of $1/N$ expansion & of $1/N$ expansion & \\
\hline

    $\pi^0\pi^0$ & $G^L_P$ & 0 & 1/2 \\

    $\pi^+\pi^-$ & $G^L_P$ & 0 & 1 \\

 $K^+K^-$ & $\sqrt\lambda\,G^L_P$ & 0 & 1 \\

$K^0\bar K^0$ & $\sqrt\lambda\,G^L_P$ & 0 & 1 \\

$\eta\eta$ & $G^L_P(\cos^2\theta+\lambda\sin^2\theta)$ &
$2G^{NL}_P\left(\cos^2\theta-\sqrt{\frac\lambda2}\sin^2\theta\right)^2$
& 1/2 \\

$\eta\eta'$ & $G^L_P(1-\lambda)\sin\theta\cos\theta$ &
$2G^{NL}_P\left(\cos\theta-\sqrt{\frac\lambda2}\sin\theta\right)$ & 1\\
&& $\times\left(\sin\theta+\sqrt{\frac\lambda2}\cos\theta\right)$ & \\

$\eta'\eta'$ & $G^L_P(\sin^2\theta+\lambda\cos^2\theta)$ &
$2G^{NL}_P\left(\sin\theta+\sqrt{\frac\lambda2}\cos\theta\right)^2$ &
1/2 \\

\hline

$\rho^0\rho^0$ & $G^L_V$ & 0 & 1/2 \\

$\rho^+\rho^-$ & $G^L_V$ & 0 & 1 \\

$K^{*+}K^{*-}$ & $\sqrt\lambda\,G^L_V$ & 0 & 1 \\

$K^{*0}\bar K^{*0}$ & $\sqrt\lambda\, G^L_V$ & 0 & 1 \\

$\omega\omega$ & $G^L_V$ & $2G^{NL}_V$ & 1/2 \\

$\omega\phi$ & $G^L_V(1-\lambda)\varphi_V$ &
$2G^{NL}_V\left(\sqrt{\frac\lambda2}+\varphi_V
\left(1-\frac\lambda2\right)\right)$ & 1 \\

$\phi\phi$ & $\lambda\,G^L_V$ & $2G^{NL}_V\left(\frac\lambda2
+\sqrt{2\lambda}\,\varphi_V\right)$ & 1/2 \\
\hline\hline
\end{tabular}
\end{table}

The decays $glueball\to two\,q\bar q$-$mesons$ may be realized through
both planar quark--gluon diagrams and non-planar ones, the contribution
from non-planar diagrams being suppressed in terms of the
 $1/N$-expansion \cite{t'hooft}. One may expect that in the
next-to-leading order the vertices are suppressed as
$G^{NL}_P/G^L_P\sim1/10$, $G^{NL}_V/G^L_V\sim1/10$ --- in any case
such a level of suppression is observed in the decay of scalar glueball
$f_0(1200-1600)$ \cite{kmatYF}. Therefore, the next-to-leading terms
are important for the channel
$glueball\to\omega\phi$ only, for other channels they may be omitted.

In the Particle Data compilation \cite{PDG} there is a narrow state
$f_J(2220)$, with $J^{PC}=2^{++}$ or $4^{++}$ and
$\Gamma\simeq23\,$MeV, which is sometimes  discussed as a candidate
for tensor glueball, under the  assumption $J=2$
(see  \cite{Doser} and references therein).  If this state does
exist with $J=2$, we see that there is no room for it on the
$q\bar q$-trajectories shown in Fig. 1: in this case it should be also
considered as an exotic state.

In the mean time there exist two statements about the value of glueball
width: according to \cite{Wein}, it should be less than hadronic widths
of $q\bar q$-mesons, while, following \cite{ufn04,PR-exotic}, it must
be considerably greater. The arguments given in \cite{Wein} are based
on the evaluation of the decay couplings in lattice calculations.
However, such calculations does not take into account the
large-distance processes, such as
$meson(1)\to real\,mesons\to meson(2)$ in case of resonance
overlapping. And just these transitions are responsible for the large
width of a state which is exotic by its origin  \cite{PR-exotic}.
The phenomenon of width accumulation for meson resonances has been
studied in \cite{APS-PL,APS-ZP,AAS-PL,AAS-ZP}, but much earlier this
phenomenon  was observed in nuclear physics
\cite{Shapiro,Okun,stodolsky}.  Therefore, I think that at present time
just the large width of $f_2(2000)$ is an argument in
favour of the glueball origin of this resonance. But to prove the
glueball nature of $f_2(1200)$ the measurement of decay constants
and their comparison to the ratios given in Table is needed.

I am grateful to L.G. Dakhno, S.S. Gershtein, V.A. Nikonov and A.V.
Sarantsev for stimulating discussions, comments and help. The paper was
supported by the Russian Foundation for Basic Research,
project no. 04-02-17091.

\newpage
\begin{figure}
\centerline{\epsfig{file=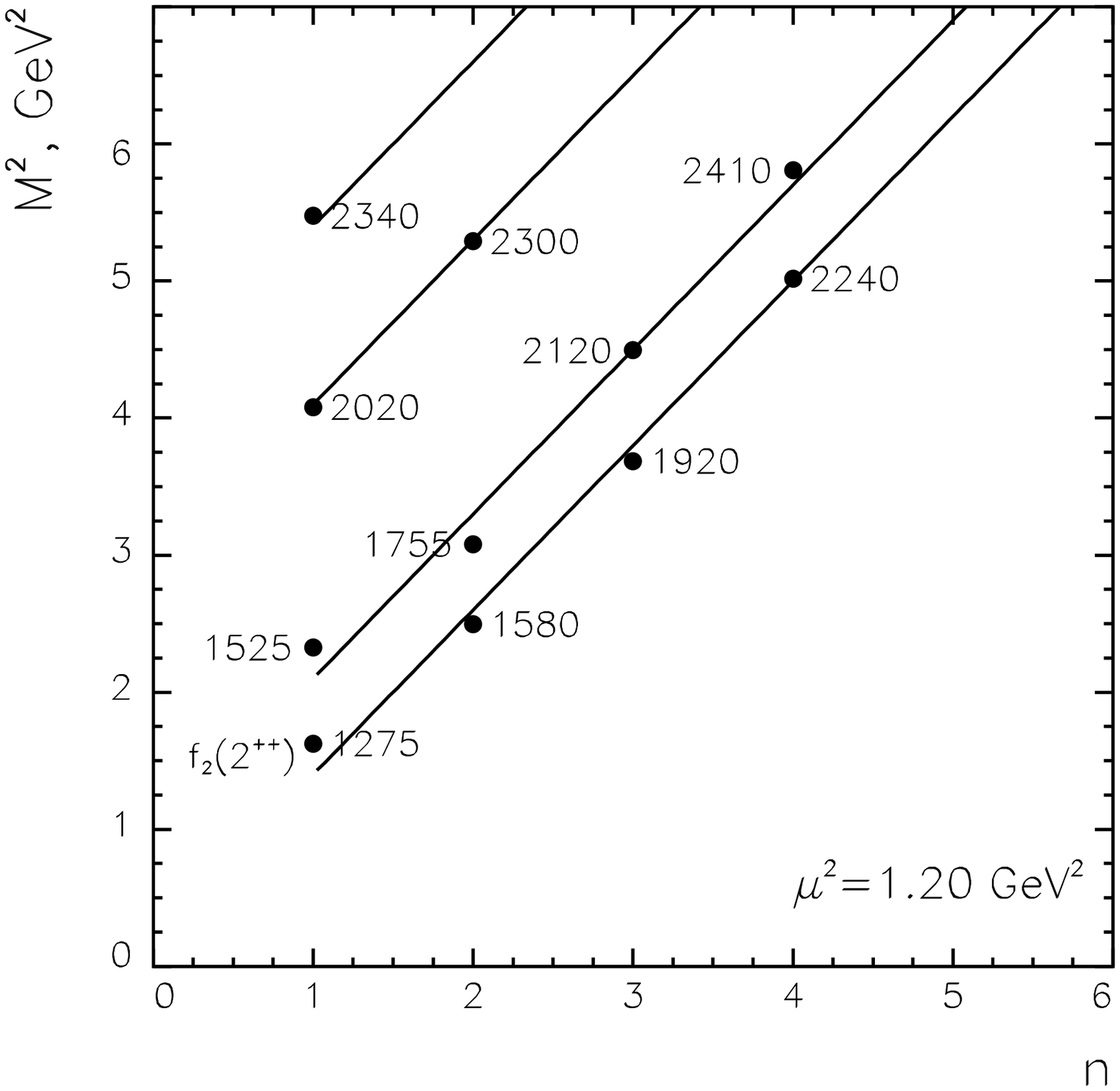,width=10cm}}
\caption{The $f_2$ trajectories of  on the $(n,M^2)$ plane; $n$ is
radial quantum number of $q\bar q$ state. The numbers stand for the
experimentally observed $f_2$-meson masses $M$.}

\end{figure}

\end{document}